\documentclass[12pt,a4paper]{article}
\pdfoutput=1
\usepackage{jheppub}
\usepackage{hyperref, amsmath, amssymb, graphicx, amsfonts, latexsym, bbold, mathbbol}
\usepackage{color}

\newcommand{\be}{\begin{equation}}
\newcommand{\ee}{\end{equation}}
\newcommand{\bea}{\begin{eqnarray}}
\newcommand{\eea}{\end{eqnarray}}
\def\eqa{&=&} 
\def\ccr{\nonumber\\} 
\newcommand{\dslash}{\partial \! \! \!  /} 
\newcommand{\Dslash}{\nabla \! \! \! \!  / \, } 
\newcommand{\bpsi}{\overline{\psi}}
\newcommand{\btheta}{\overline{\theta}}
\newcommand{\blambda}{\overline{\lambda}}

\def\la{\langle}
\def\ra{\rangle}

\author{Fiorenzo Bastianelli and Riccardo Martelli}

\affiliation{Dipartimento di Fisica e Astronomia, Universit{\`a} di Bologna and\\
INFN, Sezione di Bologna, via Irnerio 46, I-40126 Bologna, Italy}

\abstract{We calculate the trace anomaly of a Weyl fermion coupled to gravity by using Fujikawa's 
method supplemented by the choice of a consistent regulator. The latter is constructed out of Pauli-Villars regulating fields.
The motivation for presenting such a calculation stems from recent studies that suggest that the trace anomaly 
of chiral fermions in four dimensions might contain an imaginary part proportional to the Pontryagin density.
We find that the trace anomaly of a Weyl fermion is given by half  the trace anomaly of a Dirac fermion, so that
no imaginary part proportional to the Pontryagin density is seen to arise.}


\title{On the trace anomaly of a Weyl fermion}

\begin{document}
\maketitle
\flushbottom

\section{Introduction}

Trace anomalies, also called conformal or Weyl anomalies, characterize conformal field theories and find 
many applications in theoretical physics, see \cite{Duff:1993wm} for a review (and \cite{Duff:2012} for a recent update).
They consist in the fact that the trace of the  energy-momentum tensor of conformal field 
theories vanishes at the classical level,  but acquires anomalous terms at the quantum  level. 
These terms depend on the background geometry of the spacetime on which the conformal field theories are coupled to.

In a recent study \cite{Bonora:2014qla}, see also \cite{Bonora:2015nqa, Bonora:2015odi}, 
the case of chiral theories in four dimensions has been analyzed anew, considering in particular the model 
of a massless Weyl fermion. It was found that its trace anomaly contains a term proportional to the Pontryagin
density of the curved background, which appears with an imaginary coefficient. 
Such a term is indeed allowed by the consistency conditions  \cite{Bonora:1985cq, Bonora:1984ic, Boulanger:2007st},
but its emergence is nevertheless surprising. If present, it would have interesting consequences
 \cite{Bonora:2014qla, Mauro:2014eda}.

One reason to find it surprising, is that by CPT a four dimensional left handed fermion 
has a right handed antiparticle, which is expected to contribute oppositely to any chiral imbalance
as far as the coupling to gravity is concerned. Indeed, one may cast the quantum field theory of 
a Weyl fermion $\lambda$, which necessarily contains its hermitian conjugate   
$\lambda^\dagger$,  as the quantum field theory of a Majorana fermion. When coupled to gravity  
the latter gives rise to a functional determinant that can be regulated in euclidean space to keep it
manifestly real. This is achieved using, for example, massive Pauli-Villars Majorana fermions 
and  this construction excludes the appearance of a phase that might produce an imaginary term
 in the anomaly \cite{AlvarezGaume:1983ig}.

Nevertheless, formal reasonings necessitate explicit verifications, and since the results
of \cite{Bonora:2014qla} point to a different conclusion, we undertake here  the task 
of computing independently the trace anomaly of a Weyl fermion.

The calculation of trace anomalies can be performed in a variety of ways, though 
chiral theories are particularly subtle.  
Here we choose a Pauli-Villars regularization \cite{Pauli:1949zm}, employed in  \cite{AlvarezGaume:1983ig}
to compute gravitational anomalies and to make the above reasoning. More precisely, 
we cast the anomaly computation in the form of a heat kernel computation, as appearing in the method 
of Fujikawa to evaluate anomalies \cite{Fujikawa:1979ay, Fujikawa:1980vr} 
 (see also \cite{Fujikawa:2004cx, Bastianelli:2006rx}). 
Fujikawa's method recognizes the anomaly as arising form the non-invariance of the path integral measure,
and the heat kernel of a suitable positive definite differential operator (the regulator) is used to regularize and compute
the path integral jacobian that produces the anomaly. 
We supplement this  method by the scheme of ref.  \cite{Diaz:1989nx}, which allows to find a consistent regulator.
This scheme employs massive Pauli-Villars fields and makes the calculation equivalent to a one-loop Feynman 
graph regulated \`a la Pauli-Villars. This connection guarantees that by performing a Fujikawa-like 
calculation one obtains a consistent anomaly, i.e. an anomaly that satisfies the consistency conditions.
Pauli-Villars regularization has been studied in the context of trace anomalies also in \cite{Asorey:2003uf}.

In the following we first review the calculation of the trace anomaly of a Dirac field
along the lines described above. This prepares the stage for the subsequent analysis of the trace anomaly of 
a Weyl field, which appears in the following section. Eventually, 
we give  our final comments. For completeness we include some appendices.
In appendix \ref{app-A}  we report our conventions for the gamma matrices, chiral spinors, and covariant derivatives.
In appendix \ref{app-B} we review the scheme  of ref.  \cite{Diaz:1989nx} for constructing consistent regulators, 
exemplified in appendix \ref{app-C}  with the trace anomaly of a scalar field. 
 
\section{The trace anomaly of a Dirac field}

As a preliminary step, let us recall that anomalies must satisfy consistency conditions, 
integrability conditions that arise since the 
anomaly can be seen as emerging from the variation of an effective action \cite{Wess:1971yu}.
For the case of the trace anomaly due to the quantum breaking of the Weyl invariance  these
consistency conditions have been analyzed in refs. \cite{Bonora:1985cq, Bonora:1984ic, Boulanger:2007st}, 
and state that in four dimensions the trace of the stress tensor can only acquire  anomalous contributions
from the curved background proportional to
the Euler topological density $E_4$, the square of the Weyl tensor $C^2$, and the 
Pontryagin density $P_4$. A term proportional to 
 $\square R$ may also appear, but it is not universal and can be eliminated by adding a local counterterm
 to the effective action.
 The above geometrical quantities are defined explicitly by
 \bea
 E_4  \eqa R_{\mu \nu \rho \sigma} R^{\mu \nu \rho \sigma} - 4 R_{\mu \nu } R^{\mu \nu} +  R^2 \ccr
 C^2 \eqa C_{\mu \nu \rho \sigma}C^{\mu \nu \rho \sigma} = R_{\mu \nu \rho \sigma} R^{\mu \nu \rho \sigma} - 2 R_{\mu \nu } R^{\mu \nu} + \frac13 R^2  \ccr
 P_4 \eqa
  \sqrt{g} \epsilon_{\mu \nu \rho \sigma} R^{\mu \nu \alpha \beta} R^{\rho \sigma}{}_{\alpha \beta}  \;.
\eea
The topological Pontryagin density $P_4$ has never been observed to arise in conformal field theories
until the recent claims made in  \cite{Bonora:2014qla} for the case of a chiral fermion.

In this section we review the calculation of the trace anomaly of a Dirac fermion, anticipating the methods
to be used for the case of a Weyl fermion.

\subsection{Coupling to curved space and classical symmetries}

The lagrangian of a massless Dirac fermion $\psi$  in a curved spacetime reads 
\be
{\cal L} = - e\,  \bpsi \gamma^\mu \nabla_\mu \psi 
\label{D-uno}
\ee
where $e$ is the determinant of the vielbein $e^a{}_\mu$ and $\nabla_\mu$ is the covariant derivative for both
change of coordinates (diffeomorphisms) and local Lorentz transformations.
The spinor field is a scalar under diffeomorphisms and transforms as a spinor under 
local Lorentz transformations, so that  in \eqref{D-uno} the covariant derivative 
$\nabla_\mu$ contains just the spin connection
\be
\nabla_\mu = \partial_\mu + \frac14 \omega_{\mu ab} \gamma^a\gamma^b \;.
\ee
In addition, the gamma matrix  appearing  in \eqref{D-uno} contains the inverse $e^\mu{}_a$ of the vielbein, 
i.e. $\gamma^\mu= e^\mu{}_a  \gamma^a $.
Our precise conventions on spinors and covariant derivatives are found in appendix \ref{app-A}.

The Dirac action $S=\int d^4 x\, {\cal L}$ is invariant under diffeomorphisms and local Lorentz transformations.
Additional symmetries are vector and axial $U(1)$ phase transformations, and Weyl rescalings.
The axial $U(1)$ and Weyl symmetries become anomalous at the quantum level.
Let us recall here just the Weyl symmetry, which is responsible for the  energy momentum tensor 
to be traceless. It is a background symmetry 
that transforms the Dirac spinor and the background vielbein by the rules
\bea
\psi (x)  &\to&  \psi' (x)  = e^{-\frac{D-1}{2}\sigma (x)} \psi (x)  \ccr[1mm]
\bpsi (x)  &\to&  \bpsi' (x)  = e^{-\frac{D-1}{2}\sigma (x)} \bpsi (x)  \ccr[1mm]
e_\mu^a(x) &\to&  {e'}_\mu^a(x) = e^{\sigma (x)} e_\mu^a(x) 
\label{weyl-rules}
\eea
where $\sigma(x)$ is an arbitrary function.
The scaling of the fermion is easily fixed by looking at  constant scaling, then using the transformation 
properties of the spin connection one verifies the full Weyl symmetry.
We have written it in arbitrary $D$ spacetime dimensions, but we will only consider the anomaly in $D=4$.

The energy momentum tensor may be  conveniently defined by 
\be
T_{\mu a}(x) = \frac{1}{e}\frac{\delta S}{\delta e^{\mu a} (x)}
\ee
and it is covariantly conserved, symmetric, and traceless as consequence of diffeomorphisms, 
local Lorentz invariance, and Weyl symmetry, respectively.

\subsection{Consistent regulators and the trace anomaly}
 
At the quantum level one gets anomalies for the Weyl symmetry  and for the axial U(1)
phase transformations. The stress tensor acquires a trace  which depends 
on the background geometry and similarly the divergence of the axial current becomes non vanishing 
by a term proportional to the Pontryagin density of the curved background.

Let us review the trace anomaly by using the method
recalled in appendix \ref{app-B},  applied to a scalar field in appendix 
\ref{app-C} for further clarification. It consists in using a Pauli-Villars (PV) regularization
and recognizing that the anomaly comes from the mass term of the Pauli-Villars fields. 
The one-loop anomaly is then cast as a standard Fujikawa anomaly calculation, 
whose final result may be found directly by consulting the literature about heat kernels.
  
For setting our notations, let us first consider the flat space limit.
The lagrangian of a massless Dirac spinor $\psi$ reads
\be
{\cal L} = -  \bpsi \dslash \psi 
\ee
with $\dslash = \gamma^a \partial_a$.
Collecting the dynamical variables into a column vector $\phi$ as
\be
\phi = \left( \begin{array}{c} 
\psi \\ 
\bpsi^T 
\end{array} \right) 
\ee
permits to cast the lagrangian in the symmetric form (up to total derivatives)
\be 
{\cal L} = \frac12 \phi^T T {\cal O} \phi 
\label{generic-lag}
\ee
which identifies the kinetic matrix $T {\cal O}$ 
\be
T  {\cal O} =
\left( \begin{array}{cc}
  0 &  -\dslash^T \\
-\dslash  &0  
\end{array} \right) 
\ee
with $\dslash^T =\gamma^{aT}\partial_a$ containing the transposed gamma matrices.
This set up coupled to gauge fields
 was used in \cite{Diaz:1989nx} to reproduce the chiral anomaly of a Dirac fermion. 
Here we find it more convenient to use the charge conjugated field 
\be
\psi_c = C^{-1} \bpsi^T
\ee  
as independent variable rather than $\bpsi^T$, 
since  $\psi_c$ has the same index structure of $\psi$ (it lives in the same space). Here $C$ is the charge conjugation
matrix, with properties  reviewed in appendix \ref{app-A}.
Then using 
\be
\phi = \left( \begin{array}{c} 
\psi \\ 
\psi_c 
\end{array} \right) 
\ee
the free lagrangian is cast again in the form \eqref{generic-lag} with the kinetic operator given by 
\be
T  {\cal O} =
\left( \begin{array}{cc}
  0 &  C \dslash \\
C \dslash  &0  
\end{array} \right)  \;.
\ee

It may be regulated by a PV massive Dirac fermion $\theta$ with Dirac mass term and lagrangian 
\be
{\cal L} = -  \btheta \dslash \theta  - M \btheta \theta  \;.
\ee
Denoting collectively the PV fields by
\be
\chi = \left( \begin{array}{c} 
\theta \\ 
\theta_c 
\end{array} \right) 
\ee
one rewrites their lagrangian in  the form
\be 
{\cal L}_{PV} =  \frac12 \chi^T T  {\cal O} \chi +\frac12 M \chi^T T \chi
\ee 
which contains the  mass matrix $T$ and  permits the identification of the operator 
${\cal O} = T^{-1} T \cal{O}$. We find
\be
T  =
\left( \begin{array}{cc}
  0 &  C \\
C  &0  
\end{array} \right)  \;, \qquad 
{\cal O} =
\left( \begin{array}{cc}
  \dslash &0 \\
0 & \dslash    
\end{array} \right)  
\ee
out of which one constructs 
 \be
{\cal O}^2 =
\left( \begin{array}{cc}
  \dslash ^2 &0 \\
0 & \dslash^2    
\end{array} \right) 
\ee
which is negative definite in euclidean space. It is used to identify a positive definite regulator 
 ${\cal R}= - {\cal O}^2$.  Once covariantized it will enter the anomaly calculation by insertion of 
 $e^{-\frac{\cal R}{M^2}}$,  which cuts off higher frequencies, see eq. \eqref{fuj-jac}.

Thus, let us proceed with the covariantization. The lagrangian  \eqref{D-uno}
is regulated by the PV field with a Dirac mass term and covariant lagrangian  
\be
{\cal L}_{PV} = - e\,  \btheta \Dslash \theta - e M \btheta \theta 
\label{PV-uno}
\ee
with $\Dslash= e^\mu{}_a  \gamma^a  \nabla_\mu$. The kinetic term has the same invariances of the field
it regulates, but the mass term fails to be Weyl invariant and will source the trace anomaly.
From the PV lagrangian one finds 
\be
T {\cal O} =
\left( \begin{array}{cc}
  0 &  eC \Dslash \\
e C \Dslash &0  
\end{array} \right)  \;, \qquad 
T  =
\left( \begin{array}{cc}
  0 &  eC \\
eC  &0  
\end{array} \right)  \;, \qquad 
{\cal O} =
\left( \begin{array}{cc}
  \Dslash &0 \\
0 & \Dslash    
\end{array} \right)  \;.
\ee
and accordingly one identifies 
 \be
{\cal O}^2 =
\left( \begin{array}{cc}
  \Dslash ^2 &0 \\
0 & \Dslash^2    
\end{array} \right)  
\label{D-reg}
\ee
and the positive definite (in euclidean space) regulator ${\cal R}= - {\cal O}^2$.

Let us now look at the mass term that sources the trace anomaly. 
The infinitesimal  version of the Weyl transformation are found form 
\eqref{weyl-rules}.
We denote by $\delta \phi =K \phi$ and $\delta \chi =K \chi$ the infinitesimal  Weyl transformation 
on $\phi$ and $\chi$, respectively. Thus one finds that the anomalous variation of the mass terms 
may be rewritten in the form
\bea
 J \eqa K + \frac12 T^{-1} \delta T + \frac12 \frac{\delta {\cal O}}{M}  \ccr
 \eqa \frac12 \sigma + \frac12 \frac{\delta {\cal O}}{M} 
 \eea
 as described in eq. \eqref{av} of appendix \ref{app-C}.
 As discussed there this is identified with the correct Fujikawa jacobian $J$ 
 to be regulated by \eqref{D-reg}, so that  the trace anomaly is represented  by\footnote{
 By ${\cal A}$  we denote (minus) the Weyl variation of the effective action. It is related to the 
  trace of the  induced stress tensor. In euclidean space, which is where one usually performs 
  the calculations,  the precise relations are given by
  $  {\cal A} =  -  \delta_{\sigma(x)}  W[g] = \delta_{\sigma(x)}  \ln Z[g] =  
\int d^4x {\sqrt g(x)} \sigma(x) \la T^\mu{}_\mu(x) \ra$. See appendix \ref{app-C} for further details.}
\be
{\cal A} = -\lim_{M\to \infty} {\rm Tr} [ J\, e^{\frac{{\cal O}^2}{M^2}} ]
\label{a1}
\ee
where the overall minus sign (compared to eq. \eqref{wa}) is due to the fermionic nature of the fields. The limit 
$M\to \infty$ removes the regulating PV fields by making them infinitely massive
(possible negative powers in $M$ are removed by renormalization, achieved by using a set of PV fields with suitable coefficients, so that only the $M^0$ term survives the limit).
 As for the term $\delta {\cal O}$ one may compute it explicitly in $D=4$
 \be
\delta {\cal O} =
\begin{pmatrix}
- \sigma \Dslash +\frac32 (\dslash  \sigma) & 0\\
0 & - \sigma \Dslash +\frac32 (\dslash \sigma) 
\end{pmatrix}
\ee
to recognize that it will never contribute to the anomaly 
 (it only produces the Dirac trace of an odd number of gamma matrices, which vanish).
 Thus, one is left with the anomaly being obtained  by using the simplified jacobian
\be
J= \frac12 \sigma 
\ee
in \eqref{a1}.
By inspection one  verifies that the two fermions $\psi$ and $\psi_c$ contribute equivalently 
to the anomaly, so that one may 
restrict the functional trace to be just 
on the functional space of a single Dirac  spinor $\psi$, letting it contribute twice
to account for $\psi_c$. This casts the anomaly in the form
\be
{\cal A} = -\lim_{M\to \infty} {\rm Tr} [ \sigma\, e^{\frac{{\Dslash}^2}{M^2}} ]
\label{dirac-1}
\ee
where the functional trace now contains a trace over the four dimensional Dirac matrices.
This is just as the direct application of the Fujikawa method would give, but it proves that the
regulator used is indeed consistent. 

The final calculation is done in euclidean space, where the regularization is better implemented. The 
explicit result is well-known, and may be read off by consulting the literature about heat kernels, as for example
\cite{DeWitt:1965jb}. We report it here for completeness
\be
{\cal A}= \frac{1}{360 (4 \pi)^2} \int d^4 x \sqrt{g} \sigma (x) 
( 7 R_{\mu \nu \lambda \rho} R^{\mu \nu \lambda \rho} +8 R_{\mu \nu} R^{\mu \nu } -5 R^2 +12 \square R)
\ee
that translates into 
\bea
\la T^\mu{}_\mu \ra \eqa 
\frac{1}{360 (4 \pi)^2} ( 7 R_{\mu \nu \lambda \rho} R^{\mu \nu \lambda \rho} +8 R_{\mu \nu} R^{\mu \nu } -5 R^2 +12 \square R)
\ccr
\eqa \frac{1}{360(4 \pi)^2}  (-11E_4 + 18 C^2 + 12 \square R) \;.
\label{dirac-2}
\eea
The first two coefficients in the last way of casting the anomaly 
have an invariant meaning, with the first one proportional to the Euler density $E_4$
being the type A anomaly, and the second one proportional to the Weyl invariant $C^2$ being the type B anomaly,
according to the geometric classification of ref. \cite{Deser:1993yx}.
The last term is a trivial anomaly that could have been dropped.

\section{The trace anomaly of a Weyl field}

\subsection{The coupling to curved space and classical symmetries}

Let us now consider the case of a left handed Weyl spinor $\lambda$ defined by the chiral constraint
\be
\gamma^5 \lambda = \lambda 
\label{chiral-cons}
\ee
with $\gamma^5$ the chirality matrix (see  appendix \ref{app-A} for our conventions).

The coupling to gravity proceeds in the same way as for a Dirac spinor, so that  the covariant  lagrangian 
of a massless Weyl field $\lambda$  reads 
\be
{\cal L} = - e\,  \blambda \gamma^\mu \nabla_\mu \lambda \;.
\label{uno}
\ee
The action is invariant under diffeomorphisms, local Lorentz transformations, Weyl transformations
and chiral $U(1)$ phase transformation, with the same transformation rules of the Dirac field
as they preserve the chiral constraint \eqref{chiral-cons}.
In particular the Weyl symmetry has the same form appearing in eq. \eqref{weyl-rules}  and 
the energy momentum tensor of the chiral fermion remains traceless at the classical level.
It develops an anomaly at the quantum level.

It may be useful to keep in mind also the chiral $U(1)$ phase transformations 
\bea
\lambda(x) &\to&  \lambda'(x) = e^{i\alpha} \lambda(x)  \ccr[1mm] 
\blambda(x) &\to&  \blambda'(x) = e^{-i\alpha} \blambda(x) 
\label{u1}
\eea
which gives rise to the covariantly conserved axial current 
\be
J^\mu = i \blambda\gamma^\mu \lambda 
\ee
that becomes anomalous at the quantum level as well.

\subsection{Consistent regulators for the Weyl fermion and the trace anomaly}

To calculate the trace anomaly we use the method applied before. We need to introduce massive 
PV fields, which we first discuss in flat space. We must  consider left handed Weyl spinors $\theta$, defined 
by the chiral constraint 
\be
\gamma^5 \theta = \theta \;, 
\label{theta-chiral-cons}
\ee
and need them to be massive. It is not possible to utilize a Dirac mass term, as the Lorentz 
scalar $\btheta \theta$ vanishes. However one may use  a Majorana mass term of the form
\be
\Delta {\cal L}_M= \frac{M}{2} ( \theta^T C \theta + h.c.)
\ee
with $M$ a real mass parameter, $C$ the charge conjugation matrix, and $h.c.$ denoting the hermitian conjugate. 
It is real, Lorentz invariant, and nonvanishing for Grassmann valued spinors ($C$ is antisymmetric). 
This will be enough to regulate the original Weyl spinor $\lambda$.
When extended to curved space the Majorana mass violates both the Weyl symmetry \eqref{weyl-rules} and the 
chiral $U(1)$ symmetry \eqref{u1}, which therefore may both become anomalous.
To be more explicit, let us rewrite the  Majorana mass term as
 \bea
\Delta {\cal L}_M \eqa \frac{M}{2} ( \theta^T C \theta + h.c.) = 
  \frac{M}{2} ( \theta^T C \theta -\btheta C^{-1} \btheta^T) \ccr
  \eqa \frac{M}{2} ( \theta^T C \theta +\theta_c^T C \theta_c)
   \label{majorana-mass}
 \eea
 where in the last line we have used the charge conjugated field $\theta_c = C^{-1} \btheta^T$
 rather then $\btheta^T$. One may 
consider $\theta$ and $\theta_c$ as independent fields when varying the action 
to find the field equations, as well as for performing the path integral quantization.

We now turn to curved space. The covariant massive PV lagrangian is given by
\be
{\cal L}_{PV} = - e\,  \btheta \Dslash \theta + e \frac{M}{2} ( \theta^T C \theta -\btheta C^{-1} \btheta^T) 
\ee
and if we introduce again the basis
\be
\chi = \left( \begin{array}{c} 
\theta \\ 
\theta_c 
\end{array} \right) 
\ee
we may cast the lagrangian in the form
\be 
{\cal L}_{PV} =  \frac12 \chi^T T  {\cal O} \chi +\frac12 M \chi^T T \chi
\ee 
with the kinetic operator and mass matrix 
given by
\be
T  {\cal O} =
\left( \begin{array}{cc}
  0 & e C \Dslash P_R\\
e C \Dslash P_L &0  
\end{array} \right)  \;,\qquad
T =
\left( \begin{array}{cc}
  e C P_L & 0\\
0 & e C P_R     
\end{array} \right)  \;.
\ee
The projection operators $P_L = \frac{1+\gamma_5}{2}$ and $P_R = \frac{1-\gamma_5}{2}$ 
have been inserted to recall that the various operators act on the space of chiral spinors,
with $\theta$ being left handed and $\theta_c$ right handed.

At this point we would like to calculate ${\cal O}= T^{-1} T {\cal O}$, but we face the problem that the mass matrix is not invertible. However, the projectors (that are not invertible) are there to project on the correct chiral spinor space, 
and on this space the mass matrix is invertible. This is enough to recognize that
 \be
{\cal O} =
\left( \begin{array}{cc}
  0 &  \Dslash P_R\\
 \Dslash P_L &0  
\end{array} \right)  
\ee
and 
\be
{\cal O}^2 =
\left( \begin{array}{cc}
\Dslash^2 P_L & 0\\
0 & \Dslash^2 P_R 
\end{array} \right)  
\ee
with the latter that  will be used as the consistent regulator for the Weyl field.

Denoting again the infinitesimal Weyl transformation on fields by $K$, we can now represent the anomaly by
\be
{\cal A} = -\lim_{M\to \infty} {\rm Tr} \Bigl[ \Bigl( K + \frac12 T^{-1} \delta T + \frac12 \frac{\delta {\cal O}}{M} \Bigr)\, e^{\frac{{\cal O}^2} {M^2}} \Bigr] \;.
\ee
as described in the appendices \ref{app-B} and \ref{app-C}, see eq. \eqref{av}. 
Defining 
\be
P =
\left( \begin{array}{cc}
P_L & 0\\
0 & P_R  
\end{array} \right)  \;
\ee
we may forget about all the projectors in the various quantities, and rewrite the anomaly as
\bea
{\cal A} \eqa -\lim_{M\to \infty} {\rm Tr} \Bigl[ \Bigl( K + \frac12 T^{-1} \delta T + \frac12 \frac{\delta {\cal O}}{M} \Bigr) P \, e^{\frac{{\cal O}^2} {M^2}} 
\Bigr ] \ccr
\eqa -\lim_{M\to \infty} {\rm Tr} \Bigl[ \Bigl( \frac12 \sigma + \frac12 \frac{\delta {\cal O}}{M} \Bigr) P \, e^{\frac{{\Dslash}^2} {M^2}} \Bigr ] .
\eea
An explicit calculation of the term $\delta {\cal O}$ leads to
\be
\delta {\cal O} =
\begin{pmatrix}
0 & - \sigma \Dslash +\frac32 (\dslash  \sigma) \\
- \sigma \Dslash +\frac32 (\dslash \sigma) & 0
\end{pmatrix}
\ee
so that it will not contribute to the anomaly since it has no diagonal entries. Thus one is left with
\be
{\cal A} = -\lim_{M\to \infty} {\rm Tr} \Bigl[ \frac12 \sigma P \, e^{\frac{{\Dslash}^2} {M^2}} \Bigr] 
\ee
where we recall that the functional trace is on the space of fields $\phi$, which contains 
both $\lambda$ and $\lambda_c$.  
We may now restrict the trace to a single four dimensional spinor space, and rewrite the anomaly as 
\bea
{\cal A} \eqa -\lim_{M\to \infty} {\rm Tr} \Bigl[ \Bigl( \frac12 \sigma \Bigl(\frac{1+\gamma^5}{2} \Bigr) + \frac12 \sigma \Bigl(\frac{1-\gamma^5}{2} \Bigr) \Bigr)  \, e^{\frac{{\Dslash}^2} {M^2}} \Bigr] \ccr
\eqa -\lim_{M\to \infty} {\rm Tr} \Bigl[ \frac12 \sigma  \, e^{\frac{{\Dslash}^2} {M^2}} \Bigr] 
\label{wta}
\eea
where the functional trace contains a trace on the four dimensional Dirac matrices.
We see that the final result is exactly half the trace anomaly of a Dirac fermion, 
eqs. \eqref{dirac-1} and \eqref{dirac-2}. 

\section{Conclusions}

We have studied the trace anomaly of a Weyl fermion using a Pauli-Villars regularization.
The term giving the one-loop trace anomaly  has been cast in the form of a standard 
Fujikawa anomaly calculation with a regulator that is guaranteed to be consistent. 
We have found that the trace anomaly due to a Weyl fermion
is precisely half the trace anomaly of a Dirac fermion.  In particular we did not find a contribution
proportional to the Pontryagin density. The way how this comes about is made quite explicit
by our calculation. In the first line of eq. \eqref{wta}
we recognize the first term as due to the left handed Weyl field $\lambda$,
which indeed would produce by itself an imaginary contribution proportional to the Pontryagin density,
see the heat kernel formulas in \cite{Christensen:1978gi, Christensen:1978md},
but then there is a second term due to the charge conjugated field
$\lambda_c$, which has opposite chirality  and contributes oppositely to the 
to the Pontryagin density, leaving a vanishing final result.

On the other hand, these intermediate contributions to the Pontryagin density would sum up if one were to compute 
the chiral anomaly related to the chiral symmetry in \eqref{u1}, 
which indeed turns out to be proportional to the Pontryagin density.

\acknowledgments{After the completion of this work we have been discussing with
the authors of ref. \cite{Bonora:2014qla}, and in particular with Loriano Bonora,
which we thank for discussions. They maintain their point that a chiral fermion should have
a Pontryagin term in the trace anomaly. We disagree in the light
of the present paper. Still, additional studies on the subject
might be welcome to clarify further the issue from other perspectives.}

\appendix   

\section{Conventions}
\label{app-A} 

\subsection{Gamma matrices}

The Dirac matrices with flat indices $\gamma^a$ satisfy the Clifford algebra
\be
\{ \gamma^a, \gamma^b \} =2 \eta^{ab} 
\label{ca}
\ee
where the Minkowski metric $\eta_{ab}$ is mostly plus. Thus $\gamma^0$ is anti-hermitian
and the $\gamma^i$'s are hermitian (we split the index $a$ into time and space components as $a=(0,i)$).
These hermiticity properties are expressed compactly by the relation
 \be
\gamma^{a\dagger} =-\beta \gamma^a \beta
\ee
where $\beta= i\gamma^0$ is the matrix used  in the definition of $\bpsi$,
the  Dirac conjugate of the spinor $\psi$
\be 
\bpsi = \psi^\dagger \beta
\ee
which makes the product $\bpsi \psi$ a Lorentz scalar.

 The chirality matrix 
$\gamma^5$ is defined by
\be 
\gamma^5=- i \gamma^0 \gamma^1\gamma^2 \gamma^3
\ee
and satisfies
\be
 \{ \gamma^5, \gamma^a \} =0 \;, \quad  
 (\gamma^5)^2 = 1\;, \quad 
 \gamma^{5\dagger} =\gamma^5 
  \;.
   \ee
It allows to introduce the left and right chiral projectors
\be
P_L = \frac{1+\gamma_5}{2} \ , \qquad P_R = \frac{1-\gamma_5}{2} 
\ee
that split a Dirac spinor $\psi$ into its left- and right-handed components (the Weyl spinors)
\be
\psi = \lambda +\rho \;, \qquad 
\left\{ \begin{array}{l}
         \lambda=\frac{1+\gamma_5}{2} \psi  \\[2mm]
        \rho  = \frac{1-\gamma_5}{2} \psi         
       \end{array} \right. \;.
        \ee
The Weyl spinors transform irreducibly under the transformations of the Lorentz group connected to the identity
(the proper, orthochronous Lorentz group): $\lambda$ is a left-handed spinor and $\rho$
is a right-handed spinor.

The charge conjugation matrix $C$ is the matrix that relates
the gamma matrices to their transposed ones by
\be
C\gamma^{a} C^{-1} = -\gamma^{a T }  \;.
\label{C-matrix}
\ee
It is antisymmetric, as can be checked by choosing a specific representation.
It is used to define the charge conjugated field $\psi_c$
\be
\psi_c = C^{-1} \bpsi^T
\ee
in which particles and antiparticles get interchanged.
Indeed, one may check that if a Dirac spinor $\psi$ satisfies the standard Dirac equation 
coupled to a $U(1)$ gauge field by a charge $e$
\be
[\gamma^a(\partial_a -ie A_a)+ m] \psi=0
\ee
then $\psi_c$ satisfies a Dirac equation with opposite charge 
\be
[\gamma^a(\partial_a +ie A_a)+ m] \psi_c=0 \;.
\ee

Note also that a chiral spinor  $\lambda$ (say with $\gamma^5 \lambda = \lambda$) has 
its charge conjugated field 
$\lambda_c$  of opposite chirality ($\gamma^5 \lambda_c = -\lambda_c$).

A Majorana spinor can be defined as a spinor  that is equal to its charged conjugated one
\be
\psi =\psi_c
\ee
so that particles and antiparticles coincide. 
This constraint is incompatible with the chiral constraint, so that Majorana-Weyl 
fermions do not exist in 4 dimensions.

Finally, we recall that for a Weyl spinor $\lambda$ the scalar $\blambda \lambda$ vanishes, 
so that a Dirac mass term cannot be introduced. On the other hand the term
\be
 \lambda^T C \lambda
 \ee
is a Lorentz scalar, and since $C$ is antisymmetric it is
non-vanishing if the spinor is taken to be  Grassmann valued.
Thus in flat spacetime a mass term of the form 
\be
\frac{M}{2}  \lambda^T C \lambda +h.c. 
\ee with $M$ real (and $h.c.$ indicating the hermitian conjugate) is allowed: 
it is real, Lorentz invariant and non-vanishing. This is the so-called Majorana mass term. 
It violates the fermion number symmetry generated by the chiral
$U(1)$ phase transformations.

\subsection{Chiral representation of the gamma matrices}

A useful representation of the gamma matrices is the chiral one,
defined in terms of two by two  blocks by
\be
\gamma^0= -i
\left( \begin{array}{cc}
 0 & \mathbb{1} \\ 
\mathbb{1} & 0
\end{array} \right) \ , \qquad 
\gamma^i = -i  \left( \begin{array}{cc}
0 &  \sigma^i \\ 
- \sigma^i & 0
\end{array} \right) 
\ee
with $\sigma^i$ the Pauli matrices, so that 
 \be
\gamma^5=
\left( \begin{array}{cc}
\mathbb{1} & 0 \\ 
0 & - \mathbb{1}
\end{array} \right) 
\;, \qquad
\beta = i\gamma^0 = 
\left( \begin{array}{cc}
0 & \mathbb{1} \\ 
\mathbb{1} & 0
\end{array} \right) \;.
\ee
It is a useful representation as the Lorentz generators $M^{ab}$ 
and the chirality matrix $\gamma^5$ take a block diagonal form.
Indeed, the spinorial representation of the Lorentz generators $M^{ab}=\frac{1}{4} [\gamma^a,\gamma^b] = \frac12 \gamma^{ab} $
become
\be
M^{0i} = \frac12 
\left( \begin{array}{cc}
\sigma^i & 0\\ 
0&-\sigma^i  
\end{array} \right) \;, \qquad 
M^{ij} = \frac{i}{2} \epsilon^{ijk} 
\left( \begin{array}{cc}
\sigma^k & 0 \\ 
 0 & \sigma^k
\end{array} \right)  
\ee
and satisfy the algebra
\be 
[M^{ab},M^{cd}]=\eta^{bc}M^{ad} \pm {\rm 3\ terms}\;. 
\label{Lie-algebra}
\ee
In this representation the Dirac field and its chiral parts take the form
\be
\psi = 
\left( \begin{array}{c}
l \\ 
r 
\end{array} \right)  \;, \qquad
\lambda = 
\left( \begin{array}{c}
l \\ 
0
\end{array} \right)  \;, \qquad
\rho = 
\left( \begin{array}{c}
0 \\ 
r 
\end{array} \right)  
\ee
where  $l$ and $r$ are two dimensional spinors (Weyl spinors) of opposite chirality.

In the chiral representation one may take the charge conjugation matrix $C$ to be given by
\be
C = \gamma^2 \beta = -i 
\left( \begin{array}{cc}
\sigma^2 & 0\\ 
0&-\sigma^2  
\end{array} \right)  \;.
\ee
By inspection one may check that in the chiral representation the charge conjugation matrix satisfies
\be
C=-C^T=-C^{-1}=-C^\dagger =C^*\;.
\ee

\subsection{Metric, vielbein, connections}

In a minkowskian spacetime we use
a mostly plus metric $g_{\mu\nu}$. The Levi-Civita connection $\Gamma_{\mu\nu}^\lambda$ 
makes the metric covariantly constant
\be
\nabla_\lambda  g_{\mu\nu}
= \partial_\lambda g_{\mu\nu} 
-\Gamma_{\lambda\mu}^{\rho} g_{\rho \nu}
-\Gamma_{\lambda\nu}^{\rho} g_{\mu \rho} =0
\ee
and it follows that  
\bea
\Gamma_{\mu\nu}^\lambda = \frac12 g^{\lambda\rho}
(\partial_\mu g_{\nu\rho}+\partial_\nu g_{\mu\rho}-
\partial_\rho g_{\mu\nu})\ .
\eea
On vectors with upper indices the covariant derivative acts as
\be
\nabla_\mu V^\lambda =\partial_\mu V^\lambda + 
\Gamma_{\mu\nu}^\lambda\; V^\nu \;.
\ee 
We use the following conventions for the  curvature tensors
\bea
[\nabla_\mu, \nabla_\nu] V^\lambda = 
R_{\mu\nu}{}^\lambda{}_\rho V^\rho \ , \ \ \ 
R_{\mu\nu}= R_{\lambda\mu}{}^\lambda{}_\nu \  ,
\ \ \ \  R= R^\mu{}_\mu 
\eea
so that the scalar curvature of a sphere is positive.

Introducing the vielbein $e^a{}_\mu$ by setting
\be
g_{\mu\nu} = \eta_{ab} \, e^a{}_\mu\, e^b{}_\nu
\ee
one gains a new gauge symmetry: the local Lorentz transformations of tangent space.
The covariant derivative needs a corresponding connection
(the spin connection) so that for a tensor (or spinor) field $V$ (with flat indices understood) one has 
\bea
\nabla_\mu V =\partial_\mu V + \frac12 \omega_{\mu a b} M^{ab} V
\eea
where $M^{ab}$ are the generators of the Lorentz group, with Lie algebra in eq. \eqref{Lie-algebra}, 
in the representation of the field $V$.
The spin connection without torsion is defined by requiring the vielbein to be covariantly constant
\bea
\nabla_\mu  e^a{}_\nu \equiv 
\partial_\mu   e^a{}_\nu -\Gamma_{\mu\nu}^\lambda e^a{}_\lambda
+ \omega_{\mu}{}^a{}_b  e^b{}_\nu =0
\eea
which can be solved for  $\omega_{\mu}{}^{a b}$ 
by
\be
\omega_{\mu}{}^{a b} = e^{b\nu} (\Gamma_{\mu\nu}^\lambda e^a{}_\lambda -\partial_\mu   e^a{}_\nu)
\ee
or equivalently by
\bea
\omega_{\mu}{}^{a b} \eqa \frac12 e^{a \lambda }e^{b \nu}
e_{c \mu  } (\partial_\nu e^c{}_{\lambda} -\partial_\lambda e^c{}_{\nu})
\ccr
 &+&\frac12 e^{a \nu}  (\partial_\mu e^b{}_{\nu} -\partial_\nu e^b{}_{\mu})
- \frac12 e^{b \nu}  (\partial_\mu e^a{}_{\nu} -\partial_\nu e^a{}_{\mu}) \;.
\label{sc}
\eea
This last expression shows manifestly the antisymmetry under exchange of the indices $a$ and $b$.
 The spin connections transforms as a gauge field under local Lorentz transformations.

\section{Anomalies and consistent regulators}  
\label{app-B} 

The origin of anomalies in (perturbative) quantum field theory can be traced back to the fact that in the computation
of loop corrections, one has to specify a regularization scheme. The latter, in general,  does not preserve all of the 
symmetries of the classical action. After renormalizing, one can eliminate the regulating parameter
(like the momentum cut-off $\Lambda$, the $\epsilon$ parameter of dimensional regularization, or the mass $M$
of Pauli-Villars fields) and it may happen that some finite, non-symmetrical terms survive,  causing the breaking 
of those symmetries not preserved by the regularization. Still, it may happen that those terms can be removed by adding  local counterterms to the effective action, whose variation cancels the anomaly.
If this is not the case, one has a true anomaly. 

In the language of generating functionals, an anomaly  means that the effective action $\Gamma$ does not satisfy 
the Ward identity corresponding to the classical symmetry.
The part which breaks the Ward identity is identified as the ``consistent" anomaly, where consistent refers to the fact that 
the anomaly is obtained from the variation of the effective action, and thus satisfies certain integrability conditions \cite{Wess:1971yu}.

In the main text we apply the method of Fujikawa \cite{Fujikawa:1979ay, Fujikawa:1980vr}
 for computing the anomalies, improved by the scheme of  ref. \cite{Diaz:1989nx} to identify a consistent regulator. 
The latter scheme makes the anomaly calculation equivalent to a Feynman graph calculation regulated 
\`a la Pauli-Villars, so that the emerging anomaly is  necessarily consistent.

In Fujikawa's method, one recognizes the anomaly as arising from the non-invariance under a symmetry variation 
of the measure $D \phi$ of the path integral (in euclidean time) 
\be
Z=\int D \phi \ e^{-S[\phi]} \;.
\label{pi}
\ee
Let us briefly review Fujikawa's method and start by considering an infinitesimal symmetry transformation 
of the form $\delta_{\alpha}  \phi^i = \alpha f^i(\phi,\partial_\mu \phi)$, with infinitesimal constant parameter 
$\alpha$, that leaves the action invariant, i.e. $\delta_\alpha S[\phi]=0$. 
Promoting the parameter $\alpha$ to be an arbitrary function $\alpha(x)$,
one identifies the Noether current $J_\mu$ associated to the symmetry by calculating
\be
\delta_{\alpha(x)} S[\phi] 
=\int d^4x\, J^\mu  \partial_\mu \alpha(x) \;.
\label{noether}
\ee
Terms proportional to an undifferentiated $\alpha$ cannot be present, 
as for constant parameter one must recover the symmetry. 
On-shell $\delta S[\phi]=0$ for arbitrary variations 
(least action principle), and after performing an integration by parts in \eqref{noether}
one deduces that the Noether current $J^\mu$ is classically conserved 
\be
\partial_\mu  J^\mu =0 \;.
\ee
The quantum theory is defined  by the path integral, which is left invariant
by a dummy change of integration variables 
\be
\int D \phi' \ e^{-S[\phi']} = \int D \phi \ e^{-S[\phi]} \;.
\label{inv-pi}
\ee 
Let us apply this property to an infinitesimal change of the form
 \be
 \phi^i \to \phi'^i= \phi^i + \delta_{{\alpha(x)}}  \phi^i
\label{change}
 \ee
  where $\delta_{{\alpha(x)}} \phi^i$ is given by an infinitesimal  symmetry transformation with the parameter $\alpha$ 
replaced by the arbitrary function $\alpha(x)$. 
In relating the path integral  written in terms of  $\phi'^i$ to the one written in terms of $\phi^i$  
(in a condensed notation we include the space-time dependence into the index $i$),
 one may use 
 \be
 S[\phi'] =S[\phi] + \delta_{\alpha(x)} S[\phi] 
  \ee
 and consider that the path integral jacobian ${\cal  J}$ may be written as
\be
{\cal J} = {\rm Det}\, \frac{ \partial \phi'^i}{\partial \phi^j}  
= 1+ {\rm Tr}\, \frac{\partial \delta_{{\alpha(x)}}\phi^i }{ \partial \phi^j}
\equiv 1 + {\rm Tr}\, J  \;.
\label{trace}
\ee
Thus, one finds from \eqref{inv-pi} 
\be
\la {\rm Tr}\, J - \delta_{\alpha(x)} S[\phi]  \ra =0
\ee
rewritten with an integration by parts as
\be
\int d^4x\,\alpha(x) \partial_\mu  \la J^\mu  \ra =  - {\rm Tr}\, J  \;.
\label{fuj-anomaly}
\ee
This shows that the Noether current is not conserved at the quantum level
if the path integral measure carries a nontrivial jacobian under \eqref{change}
\be
\partial_\mu  \la J^\mu  \ra \neq  0  \;.
\label{fuj-anomaly-2}
\ee
We have indicated by $\la  \cdots \ra$ the quantum expectation values
defined as normalized averages within the path integral. Also, 
we have assumed that the jacobian is independent of the quantum fields, so 
to pull it out of the expectation value. 

To proceed further, one must define carefully the formal expressions appearing in the above reasonings.
Ideally, one would like to fully specify the path integration measure,
so that the evaluation of the jacobian would be a well-defined task.
In practice, one is able to compute gaussian path integrals only, 
and resort to perturbative methods for more complicated  cases.
Nevertheless, one can still obtain the one-loop anomalies by regulating the 
trace in \eqref{fuj-anomaly}, as shown by Fujikawa   \cite{Fujikawa:1979ay,Fujikawa:1980vr}.
Employing a positive definite operator  ${\cal R}$
the candidate anomaly is regulated as
\be
{\rm Tr}\, J \ \ \to \ \  \lim_{M \to \infty}  {\rm Tr}\, J\, e^{-\frac{\cal R}{M^2}} \;.
\label{fuj-jac}
\ee
The functional trace is written in a more explicit notation (for a single scalar field) as
\be
{\rm Tr}\, J  = \int d^4x \int d^4y\, J(x,y) \delta^4(x-y) \;, \qquad  
J(x,y) =\frac{\delta (\delta_{\alpha(x)} \phi(x))}{\delta \phi(y)}
\ee
and regulated  by the differential operator ${\cal R}(x)$ acting on the $x$ coordinates as
\be
\lim_{M \to \infty}  {\rm Tr}\, J\, e^{-\frac{\cal R}{M^2}} =
\lim_{M \to \infty}  \int d^4x \int d^4y\, J(x,y) \, e^{-\frac{{\cal R}(x)}{M^2}} \,
 \delta^4(x-y) \;.
 \ee

For an arbitrary regulator ${\cal R}$, it is not obvious which kind of anomaly one is going to get.
A well-defined algorithm for determining those regulators ${\cal R}$ which produce consistent anomalies
 has been established in \cite{Diaz:1989nx} (see also \cite{Hatsuda:1989qy}).
The basic idea is to first use a Pauli-Villars (PV) regularization \cite{Pauli:1949zm}, compute
the anomalies due to the non-invariance of the PV mass term, and then read off the regulators 
and jacobians to be used in  the Fujikawa's scheme in order to reproduce the anomalies.
Since the PV method yields \mbox{consistent} anomalies, being a Feynman graph calculation,
one obtains consistent regulators. 

In more details the PV method for computing one-loop anomalies goes as follows. Let us denote by 
 $\phi$ a collection of quantum fields with quadratic action 
 \be 
{\cal L}_\phi = \frac12 \phi^T T {\cal O} \phi 
\label{act}
\ee
invariant under a linear symmetry of the form
\be
\delta \phi = K\phi \;.
\ee
The case of linear symmetries is enough for our purposes.
The one-loop effective action of this theory is regulated by subtracting a loop  of a massive
PV fields $\chi$ with action 
\be 
{\cal L}_\chi =  \frac12 \chi^T T  {\cal O} \chi +\frac12 M \chi^T T \chi
\label{PV-act}
\ee 
where the last term describes the mass of the PV fields\footnote{More generally, 
one should employ a set of PV fields with mass $M_i$ and weight $c_i$ to be able to regulate and cancel 
all possible one-loop divergences \cite{Pauli:1949zm}.
For the sake of the present exposition it is enough to consider 
only one copy of the PV fields. Also, the mass $M$ in the PV lagrangian generically should carry 
an appropriate positive power, according to the mass dimension of the differential operator $\cal O$.}. 
The invariance of the original action
is extended to an invariance of the massless part of the PV action by defining 
\be
\delta \chi = K \chi
\label{PV-transf}
\ee
so that only the PV mass term may break the symmetry
(if one can find a symmetrical mass term, then the symmetry will be anomaly  free).
One refers to $T {\cal O}$ as the  kinetic matrix and to $T$ as  the mass matrix.
They both depend  on eventual background fields, which may get transformed
under the symmetry variation as well.
The anomalous response of the path integral under a symmetry 
is now due to the mass term only, since the measure of the PV fields $\chi$
can be defined in such a way that their jacobian cancels the jacobian
of the original fields $\phi$, as argued in \cite{Diaz:1989nx}. Under the symmetry transformation
\eqref{PV-transf} the mass term of the PV lagrangian  varies as
\be
\delta{\cal L}_\chi =  \frac12 M \chi^T (T K +K^T T +\delta T) \chi \;.
\ee
Using this into the variation of the PV-regulated path integral one computes the anomaly 
in the Noether current as 
\bea
\int d^4x\,\alpha(x) \partial_\mu  \la J^\mu  \ra \eqa
- \lim_{M \to \infty}  
{\rm Tr} \biggl [ \frac12 M
 \biggl ( TK + K^TT + \delta T \biggr ) \biggl (TM + T {\cal O}
\biggr )^{-1} \biggr ] 
\ccr
\eqa
- \lim_{M \to \infty}  
{\rm Tr} \biggl [\biggl (K + \frac12 T^{-1} \delta T \biggr ) \biggl ( 
1+ \frac{\cal O}{M} \biggr )^{-1} \biggr ] 
\label{anomaly}
\eea
where we replaced $K^T T $ by $TK$, since both $T$ and $T {\cal O}$ are  symmetric,
and used the $\chi$-propagator from \eqref{PV-act} to close the $\chi$-loop 
(recall its relative minus sign with respect
to the $\phi$-loop). The limit  ${M \to \infty}$ 
indicates that the PV fields are removed by
making them infinitely massive, so that in \eqref{anomaly} only a mass independent 
term survives, which gives the anomaly\footnote{Eventual diverging terms are removed by using a set of PV fields 
entering the loop with suitable coefficients $c_i$, instead of a single PV field, as reminded in the previous footnote. 
It is not necessary to explicitate this procedure further. \label{foot3}}.

At this stage one may notice that the expansion of $(1 + \frac{\cal O}{M})^{-1}$ leads to Feynman graphs,
just as the expansion of $e^{-\frac{\cal O}{M}}$. Hence one may
cast the anomaly calculation as a typical calculation of a  Fujikawa's jacobian, eq. \eqref{fuj-jac},
by identifying
\be
J = K + \frac12 T^{-1} \delta T\;,
\qquad  \ {\cal R} = {\cal O} \;. 
\label{ide}
\ee 
This freedom in regulating path integral jacobians by using suitable functions of the regulator ${\cal R}$
was already noticed in  \cite{Fujikawa:1979ay,Fujikawa:1980vr},
and used in \cite{Diaz:1989nx} to make the above connection.

For many cases the regulator ${\cal O}$ is enough, while in other  cases 
(typically when ${\cal O}$ is a first order differential operator) one has to improve it. 
A way of doing this is achieved by inserting the identity $ 1 = (1 - \frac{\cal O}{M}) (1 - \frac{\cal O}{M})^{-1}$ 
into \eqref{anomaly}, so that the functional trace becomes
\bea
{\rm Tr} \biggl [\biggl (K + \frac12 T^{-1} \delta T \biggr ) \biggl ( 1- \frac{\cal O}{M} \biggr )  
\biggl ( 1- \frac{{\cal O}^2}{M^2} \biggr )^{-1} 
\biggr ] \;, 
\eea
and one finds that the jacobian and regulator to be used in a Fujikawa-like calculation
are given by 
\be
 J = \biggl (K + \frac12 T^{-1} \delta T \biggr ) \biggl ( 1- \frac{\cal O}{M} \biggr )  
 \;, 
\qquad {\cal R} = - {\cal O}^2   \;.
\label{impr-reg}
\ee
This form  of the jacobian can be simplified in many cases, with the term proportional to $\cal O$
often not contributing to the anomaly.

The construction is easily extended to fermionic systems by taking care of signs. 
In the main text it is applied to the trace anomaly of a Weyl fermion. 
To familiarize further with the PV method, we review in the next section 
its application  to the trace anomaly of a scalar field in a curved background.

\section{The trace anomaly of a scalar field}
\label{app-C} 

The action of a massless scalar field in $D$ dimensions, conformally coupled to gravity, is given by
\bea
S[\phi; g_{\mu\nu}] 
= \int d^Dx \sqrt{g}\, \frac12 
(g^{\mu\nu} \partial_\mu\phi \partial_\nu\phi +\xi R\phi^2 )
\label{uno-sca}
\eea
with $\xi = \frac{(D-2)}{4(D-1)}$. It is invariant under local Weyl rescalings of the metric,
with infinitesimal transformations given by
\bea
\delta_{\sigma(x)} 
\phi(x) \eqa \frac{2-D}{2} \sigma (x) \phi (x)
\ccr
\delta_{\sigma(x)} 
  g_{\mu\nu}(x) \eqa 2 \sigma (x) g_{\mu\nu}(x) 
  \label{Weyl-transf}
\eea
where $\sigma(x)$ is an infinitesimal arbitrary function. This symmetry implies that the energy-momentum
tensor, defined by $T_{\mu\nu} = \frac{2}{\sqrt g} \frac{\delta S}{\delta g^{\mu\nu}}$,
is traceless. This follows by considering an infinitesimal Weyl transformation which, being a symmetry,
leaves the action unchanged
\be
\delta_{\sigma(x)}  S[\phi; g_{\mu\nu}] = \int d^Dx\ \biggl
( \frac{\delta S}{{\delta g_{\mu\nu}(x)}}
\delta_{\sigma(x)}  g_{\mu\nu}(x) + \frac{\delta S}{\delta \phi (x)} \delta_{\sigma(x)}  \phi (x) \biggr ) = 0 \;.
\ee
On the shell of the $\phi$  equations of motion the second term 
vanishes by itself ($\frac{\delta S}{\delta \phi}=0 $), 
and one finds that the energy-momentum tensor is traceless
\bea
\delta_{\sigma(x)}  S[\phi; g_{\mu\nu}] \Big |_{on-shell}\eqa - \int d^Dx {\sqrt g}\, \sigma(x) T^\mu{}_\mu(x) =0 \ccr[2mm]
&\longrightarrow & \quad 
T^\mu{}_\mu(x) =0\;.
\eea
where of course $T^\mu{}_\mu(x) = g^{\mu\nu}(x) T_{\mu\nu}(x) $ is the trace of the energy-momentum tensor.
The quantum theory is defined by the path integral 
\be
Z[g] \equiv e^{-W[g]}= \int D \phi\, e^{-S[\phi; g_{\mu\nu}] }  
\ee
which produces the one-loop effective action $ W[g]$.
The Weyl variation of the effective action is proportional to the one-point function of  $T^\mu{}_\mu(x)$,
the trace of the energy-momentum tensor.
By a change of variables  it is related to the (anomalous) variation of the path integral measure under
Weyl transformations
\bea
\delta_{\sigma(x)}  W[g] \eqa - \delta_{\sigma(x)}  \ln Z[g] =  \la
\delta^{(g)}_{\sigma(x)}  S[\phi; g_{\mu\nu}] \ra \ccr
\eqa  - \int d^Dx {\sqrt g(x)} \sigma(x) \la T^\mu{}_\mu(x) \ra = -
{\rm Tr}\, \frac{\delta( \delta_{{\sigma(x)}}\phi(x))}{ \delta \phi(y)}
\eea
where the Weyl variation $\delta^{(g)}_{\sigma(x)}$ acts only on the metric $g_{\mu\nu}(x)$.
The last line shows that if the functional trace does not vanish, one finds a trace anomaly
\be
\la T^\mu{}_\mu(x) \ra \neq 0 \;.
\ee

In the original Fujikawa calculation \cite{Fujikawa:1980vr},
 it was argued that the preferred integration variables to carry out the calculation 
were given by $\tilde \phi = g^\frac14 \phi$, to guarantee that the diffeomorphisms 
would not be anomalous.
These variables had been used previously by Hawking \cite{Hawking:1976ja}.
The infinitesimal Weyl variation of the variables $\tilde \phi$ 
is fixed by \eqref{Weyl-transf} and reads
\be
\delta_{\sigma(x)}\tilde\phi(x)  = \sigma(x)\tilde\phi(x)   
\ee
so that the trace anomaly is obtained by computing
\bea
{\cal A}\equiv
\int d^Dx {\sqrt g(x)}  \sigma(x) \la T^\mu{}_\mu(x) \ra \eqa  
{\rm Tr}\,[ \sigma(x) \delta^D(x-y)] \ccr
\eqa 
\lim_{M \to \infty}  {\rm Tr}\, [ \sigma(x)  e^{-\frac{{\cal R}(x)}{M^2}} \delta^D(x-y)] 
\label{wa}
\eea
which has been regulated by using the positive definite kinetic operator of the $\tilde \phi$ field
\be
{\cal R} = -g^\frac14 \Box g^{-\frac14} + \xi R 
\ee
where $\Box = \frac{1}{\sqrt{g}} \partial_\mu {\sqrt{g}} g^{\mu\nu} \partial_\nu$ is the covariant scalar  laplacian.
One recognizes in \eqref{wa} the appearance of the heat kernel 
$ {\rm Tr}\, [ \sigma  e^{-\frac{\cal R}{M^2}}]$ of the operator $\cal{R}$,
which is known to have an expansion of the form
\be
{\rm Tr}\, [ \sigma  e^{-\frac{\cal R}{M^2}}] = \frac{M^D}{(4\pi)^\frac{D}{2}}
\int d^Dx\sqrt{g(x)} \sigma(x) \sum_{n=0}^\infty \frac{a_n(x)}{M^{2n}}
\label{hke}
\ee
where the coefficients $a_n(x)$ are called Seeley-DeWitt coefficients at coinciding points \cite{DeWitt:1965jb}.
Of course in \eqref{wa} 
the negative powers in $M^2$  vanish in the limit $M\to \infty$, while the (diverging)
positive powers in $M^2$ are eliminated by renormalization (see footnote \ref{foot3}), 
so that one is left with a trace anomaly given by the $M^0$
term present in the heat kernel expansion 
\be
\la T^\mu{}_\mu(x) \ra = \frac{1}{(4\pi)^\frac{D}{2}}\, a_\frac{D}{2}(x) \;.
\ee
As a side result of this formula one recognizes that 
in odd dimensions there are no anomalies, as the corresponding Seeley-DeWitt coefficient vanish
(it could be present if the odd dimensional space had a boundary).

The calculation of the Seeley-DeWitt coefficients, and of the trace anomalies in particular, 
can be performed in a variety of ways. One may use a similarity transformation to
the basis of the scalar field $\phi$ and corresponding scalar regulator
${\cal R} =  \Box - \xi R$ to evaluate the heat kernel expansion perturbatively with covariant methods 
\cite{DeWitt:1965jb} (see also \cite{Vassilevich:2003xt} for a review).
Another useful method to compute the anomaly
is to represent the Fujikawa trace as a quantum mechanical trace for a particle moving in curved space
\cite{Bastianelli:1991be, Bastianelli:1992ct, Bastianelli:2000dw} (see also the book  \cite{Bastianelli:2006rx}
 for details on this method). At the end the explicit results are as follows.
In $D=2$ one must set $\xi=0$ and obtains the trace anomaly
\be
\la T^\mu{}_\mu(x) \ra 
= \frac{R}{24\pi} \;. 
\ee
In $D=4$ the conformal coupling is $\xi=\frac16$ and one finds the trace anomaly
\bea
\la T^\mu{}_\mu(x) \ra 
\eqa \frac{1}{180(4\pi)^2} (R_{\mu\nu\lambda\rho} R^{\mu\nu\lambda\rho} 
- R_{\mu\nu}R^{\mu\nu} + \Box R ) \ccr
\eqa \frac{1}{360(4\pi)^2} ( -E_4+ 3 C + 2 \Box R ) \;.
\eea
  In $D=6$  one must use $\xi=\frac15$ and obtains an expression of the anomaly that is somewhat lengthy. 
Its explicit  form may be found in \cite{Bastianelli:2000hi,Bastianelli:2000dw, Bastianelli:2001tb}. 
These results for a real scalar field are well-established, of course, 
 and they satisfy the consistency conditions discussed
in refs.  \cite{Bonora:1985cq, Bonora:1984ic, Boulanger:2007st}
(see also \cite{Bonora:1983ff, Bastianelli:2000rs}).

Let us now verify that the above regulator, used in the Fujikawa's approach, is  indeed guaranteed to 
be consistent by the PV method. One may take the PV field $\chi$ 
to have a covariant mass term, so that general coordinate invariance
remains free from anomalies. However, it breaks Weyl invariance, which may thus develop an anomaly.
The PV action is given by
\bea
S[\chi;g] 
= \int d^Dx \sqrt{g}\, \frac12 
(g^{\mu\nu} \partial_\mu\chi \partial_\nu\chi +\xi R\chi^2 +M^2\chi^2 )
\label{due-sca}
\eea
rewritten with an integration by part as
\bea
S[\chi;g] 
= \int d^Dx \sqrt{g}\, \frac12   \Big ( \chi (-\Box +\xi R)\chi  + M^2 \chi^2 \Big)  \;.
\label{tre-sca}
\eea
Following the general prescriptions in \eqref{ide}, and using
the $\chi$ basis (the field $\chi$ is a scalar), one recognizes 
\be
T {\cal O}   = \sqrt{g} (-\Box +\xi R) \;, \qquad  T = \sqrt{g} 
\ee
which fixes  the jacobian 
\be
J = K + \frac12 T^{-1} \delta T =\sigma 
\ee 
and the regulator
\be 
{\cal O}   = -\Box +\xi R 
\ee
that should be used in a Fujikawa-like calculation. 

As a check one verifies that the same result is obtained
by using the Fujikawa-Hawking variable $\tilde \chi = g^\frac14 \chi$
(the field $\tilde \chi$ is now a density), which has the property of making 
the mass matrix $T$ constant (background field independent).
One reads off 
 \be
T {\cal O}   =  g^\frac14 (-\Box +\xi R) g^{-\frac14}\;,  \qquad
T = 1 \;,  
\ee
and since $\delta T =0$ one finds directly the jacobian 
\be
J = K  =\sigma 
\ee 
 and regulator 
\be
{\cal O}   = g^\frac14 (-\Box +\xi R) g^{-\frac14}\;. 
\ee
The anomaly is the same as before. Indeed the traces in the $\chi$ basis and in the $\tilde \chi$ basis 
carry different powers of $g$, which just corresponds to the  similarity transformation
that relates the two basis of the same functional space. Said differently, the scalar product in the two basis takes
the form 
\be
\la \chi_1|\chi_2 \ra = \int d^Dx \sqrt{g}\,  \chi_1^* \chi_2  = 
 \int d^Dx\, \tilde  \chi_1^* \tilde   \chi_2 \;. 
 \ee

In the main text we use the PV method to get consistent regulators for Dirac and Weyl fermions.
In those cases one finds a first order differential operator as kinetic term, so that one should improve 
the construction of consistent regulators and jacobians, as described at the end of the previous section 
in eq. \eqref{impr-reg}. By using the background invariance of the kinetic term (that is 
$\phi^T(T {\cal O} K +  \frac12 \delta(T {\cal O}) )\phi =0$),
one may cast  the jacobian $J$ in the equivalent form \cite{Hatsuda:1989qy}
\be
 J =K + \frac12 T^{-1} \delta T + \frac12 \frac{\delta {\cal O}}{M} \;, 
\qquad {\cal R} =  - {\cal O}^2   
\label{av}
\ee
which is used in the main text for a Fujikawa-like calculation of the trace anomalies of Dirac and Weyl fermions.
In this form the method had been used in \cite{Bastianelli:1990ev} 
to compute the gravitational anomaly of a chiral boson in two dimensions, 
which also carries a first order differential operator as kinetic term. 


\end{document}